\documentclass[11pt,twoside]{article}

\usepackage{asp2004}
\usepackage{epsf}
\usepackage{psfig}
\usepackage{lscape}

\begin{document}
\title{Identifying the Obscured Black-Hole Growth Phase of Distant Massive Galaxies}
\author{D.~M.~Alexander}   
\affil{Department of Physics, Durham University, Durham, DH1 3LE, UK}    

\begin{abstract} 
It is well established that a dominant phase in the growth of massive
galaxies occurred at high redshift and was heavily obscured by gas and
dust. Many studies have explored the stellar growth of massive
galaxies but few have combined these constraints with the growth of
the supermassive black hole (SMBH; i.e.,\ identified as AGN
activity). In this brief contribution we highlight our work aimed at
identifying AGNs in $z\approx$~2 luminous dust-obscured
galaxies. Using both sensitive X-ray and infrared (IR)--submillimeter
(submm) observations, we show that AGN activity is common in
$z\approx$~2 dust-obscured systems. With a variety of techniques we
have found that the majority of the AGN activity is heavily obscured,
and construct diagnostics based on X-ray--IR data to identify some of
the most heavily obscured AGNs in the Universe (i.e.,\ AGNs obscured
by Compton-thick material; $N_{\rm H}>1.5\times10^{24}$~cm$^{-2}$). On
the basis of these techniques we show that SMBH growth was typically
heavily obscured ($N_{\rm H}\ge10^{23}$~cm$^{-2}$) at $z\approx$~2,
and find that the growth of the SMBH and spheroid was closely
connected, even in the most rapidly evolving systems.

\end{abstract}


\section{Introduction}   

There is no doubt that Active Galactic Nuclei (AGN) are an important
component in the formation and evolution of galaxies. The seminal
discovery that massive galaxies in the local Universe harbor a
super-massive black hole (SMBH) indicates that they must have hosted
AGN activity over the past $\approx$~13~Gyrs (e.g.,\ Soltan 1982;
Kormendy \& Richstone 1995). Furthermore, the close relationship
between the mass of the SMBH and the spheroid in local galaxies points
towards a close connection between the growth of the SMBH and the host
galaxy (e.g.,\ Magorrian et~al. 1998; Ferrarese \& Merritt
2000). Finally, the most successful galaxy formation and large-scale
structure models require AGN outflows to suppress star formation and
reproduce the properties of massive galaxies in the local Universe
(e.g.,\ Croton et~al. 2006; Bower et~al. 2006).

Stellar population synthesis modelling of nearby galaxies and studies
of the cosmic history of star formation indicate that the bulk of the
stellar build up of massive galaxies must have occurred at high
redshift ($z\approx$~2; i.e.,\ when the Universe was $\approx$~25\%
the current age; e.g.,\ Heavens et~al. 2004; Hopkins
et~al. 2006). Under the assumption that the growth of the SMBH is
concordant with that of the galaxy spheroid, the dominant growth phase
of SMBH growth in massive galaxies must also have occurred at high
redshift. Currently the most {\it efficient} identification of AGNs
(and therefore SMBH growth) is made with deep X-ray surveys (e.g.,\
Brandt \& Hasinger 2005). For example, the deepest X-ray surveys
(e.g.,\ Alexander et~al. 2003; Luo et~al. 2008) are able to detect
even moderately luminous AGN activity at $z\approx$~2 ($L_{\rm
X}\approx10^{42}$--$10^{43}$~erg~s$^{-1}$; i.e.,\ $>10$ times below
the canonical threshold assumed for quasars). Furthermore, the high
rest-frame energies probed by these surveys at $z\approx$~2
($\approx$~1.5--24~keV) means that the selection of AGNs is {\it
almost} obscuration independent. This latter point is important since
the dominant growth phase of massive galaxies appears to have been
heavily obscured by dust and gas (e.g.,\ Chapman et~al. 2005; Le
Floc'h et~al. 2005).

In this brief contribution we hightlight some of our recent work aimed
at identifying obscured AGN activity in distant rapidly evolving
galaxies at $z\approx$~2; see \S2 \& \S3. We discuss these results
within the context of the SMBH--sperhoid growth phase of today's
massive galaxies ($M_{\rm GAL}\approx10^{11}$~$M_{\odot}$); see \S4.


\section{Heavily Obscured Black-Hole Growth in the Most Luminous $z\approx$~2 Galaxies}

 \setcounter{figure}{0}
 \begin{figure}[!t]
 \plotone{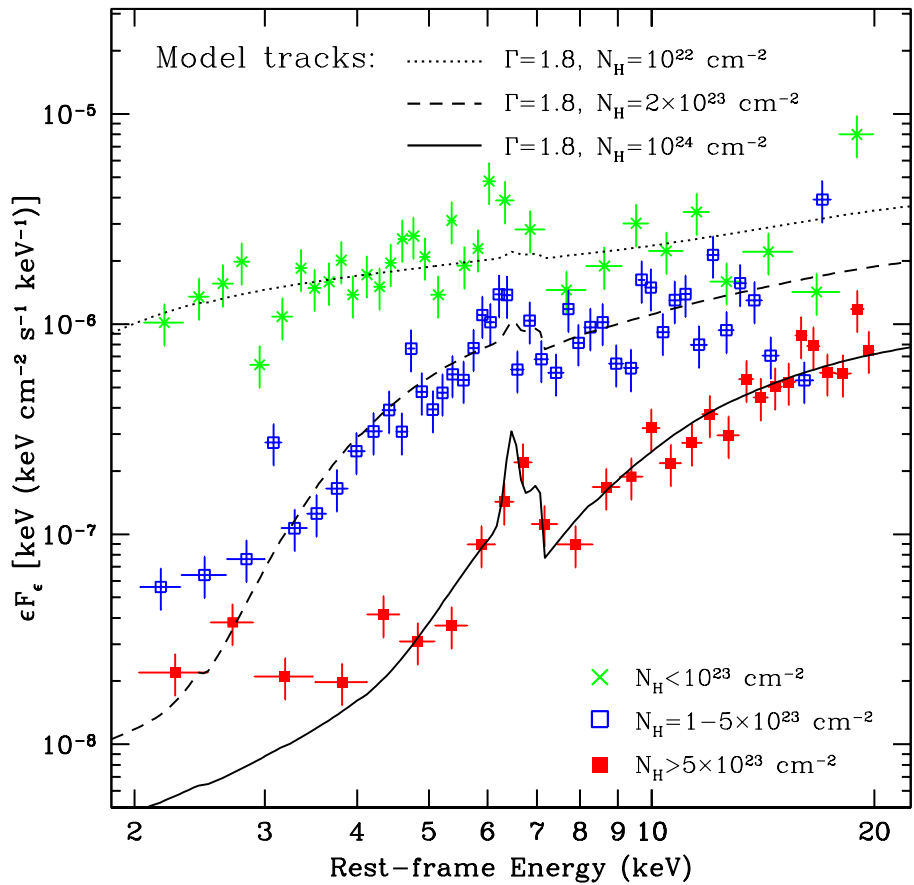}
 \caption{Composite rest-frame 2--20~keV spectra for the $z\approx$~2 SMGs hosting AGN activity in Alexander et~al. (2005b). AGNs with comparable amounts of X-ray absorption have been stacked in rest-frame energy space; as indicated. The majority ($\approx$~85\%) of the AGNs are heavily obscured ($N_{\rm H}\ge10^{23}$~cm$^{-2}$) and some may be Compton thick ($N_{\rm H}>1.5\times10^{24}$~cm$^{-2}$). See Fig.~7 of Alexander et~al. (2005b) for more details.}
 \end{figure}

Due to the negative $K$-correction for IR-luminous galaxies at $z>1$,
submm/mm surveys select the most bolometrically luminous systems in
the Universe (e.g.,\ Blain et~al. 2002; $L_{\rm
BOL}\approx10^{13}$~$L_{\odot}$). After intense multi-wavelength
follow-up observations it is now clear that submm-emitting galaxies
(SMGs; $f_{\rm 850\mu m}>4$~mJy) are gas-rich massive galaxies at
$z\approx$~2 hosting energetic star-formation activity (e.g.,\
\citealt{smail02}; \citealt{chap05}); the stellar--dynamical and CO
gas masses of these galaxies are typically
$\approx10^{11}$~$M_{\odot}$ and $\approx3\times10^{10}$~$M_{\odot}$,
respectively (e.g.,\ \citealt{swin04}; \citealt{borys05};
\citealt{greve05}). The compactness and high inferred gas density of
the CO emission from SMGs suggests that the CO dynamics trace the mass
of the galaxy spheroid (e.g.,\ \citealt{bouche07}). It is possible
that every massive galaxy ($>$~1--3~$L_{\rm *}$) in the local Universe
underwent at least one submm-bright phase at some time in the distant
past (e.g.,\ \citealt{swin06}). It is therefore interesting to ask the
question: do these intense starbursts also host AGN activity?

Early comparisons between X-ray and submm surveys showed little
overlap between the two populations. However, using the deepest X-ray
observations available (the 2~Ms {\it Chandra} Deep Field-North
survey; Alexander et~al. 2003), we found that a large fraction
($\approx$~28--50\%) of SMGs host X-ray weak AGN activity (Alexander
et~al. 2005a). With X-ray color-color diagnostics the amount of
absorption towards the AGN in individual SMGs could be assessed. By
stacking the X-ray data of AGNs with the same apparent amount of
absorption, it was then possible to construct good-quality composite
X-ray spectra, allowing the definitive features of X-ray absorption to
be identified (photo-electric absorption cut off; possible
Fe~K$\alpha$ emission); see Fig.~1. These analyses indicate that the
majority ($\approx$~85\%) of the AGNs in SMGs are heavily obscured
($N_{\rm H}\ge10^{23}$~cm$^{-2}$) and some may be Compton thick
($N_{\rm H}>1.5\times10^{24}$~cm$^{-2}$); see \S3 for other
constraints on Compton-thick AGNs. However, even correcting for the
presence of absorption, the AGN activity in SMGs was typically found
to be of moderate luminosity ($L_{\rm
X}\approx$~10$^{43}$--10$^{44}$~erg~s$^{-1}$). Given the huge bolometric
luminosities of these systems, we suggested that, although AGN
activity is often present, star formation is likely to dominate the
energetics of SMGs. {\it Spitzer}-IRS observations have indeed shown
that the mid-IR spectra of SMGs are dominated by star-formation
activity, but often have an underlying continuum component of AGN
activity (e.g.,\ Menendez-Delmestre et~al. 2007; Pope et~al. 2008).

\section{Identification of a Large Population of Compton-thick Quasars at $z\approx$~2}

The majority of the AGNs identified in the SMG population are
obscured. However, even with a 2~Ms {\it Chandra} exposure, the most
heavily obscured AGNs can be weak or undetected at X-ray energies. 
The most challenging sources to identify are Compton-thick
AGNs with $N_{\rm H}>1.5\times10^{24}$~cm$^{-2}$. In the local
Universe, Compton-thick AGNs comprise $\approx$~50\% of the AGN
population (e.g.,\ Risaliti et~al. 1999) and, given claims of a
possible increase in AGN obscuration with redshift (e.g.,\ La Franca
et~al. 2005), it is expected that at least $\approx$~50\% of the
distant AGN population will also be Compton thick. The most robust
identification of Compton-thick AGNs is made with X-ray spectroscopy,
where the presence of a high equivalent width Fe~K emission line and a
steeply rising reflection component at $>10$~keV reveals that little
or no X-ray emission is being seen directly (e.g.,\ George \& Fabian
1991). However, the extreme absorption towards a Compton-thick AGN
renders the observed X-ray emission orders of magnitude fainter than
the intrinsic emission (i.e.,\ the emission corrected for
absorption). For example, the nearby well-studied AGN NGC~1068 has an
intrinsic X-ray luminosity comparable to AGNs identified in SMGs but,
due to the presence of extreme absorption ($N_{\rm
H}\approx10^{25}$~cm$^{-2}$; i.e.,\ heavily Compton thick), if placed
at $z\approx$~2 it would have an estimated flux $\approx$~10 times
below the sensitivity limit of the 2~Ms CDF-N observations. This
illustrates the considerable challenges in identifying distant
Compton-thick AGNs.\footnote{Indeed, only $\approx$~50 Compton-thick
AGNs have been robustly identified in the Universe, most at low
redshift, a paltry $\approx10^{-9}$ of the cosmic population (Comastri
2004)!}  Clearly, in order to be able to provide a complete census of
AGN activity (and therefore the cosmic history of SMBH growth) it is
necessary to be able to identify such heavily obscured AGNs without
the need for high-quality X-ray spectroscopy; the latter will be
challenging before the next generation of X-ray observatories,
requiring $\gg10$~Ms X-ray exposures.

 \setcounter{figure}{1}
 \begin{figure}[!t]
 \plotone{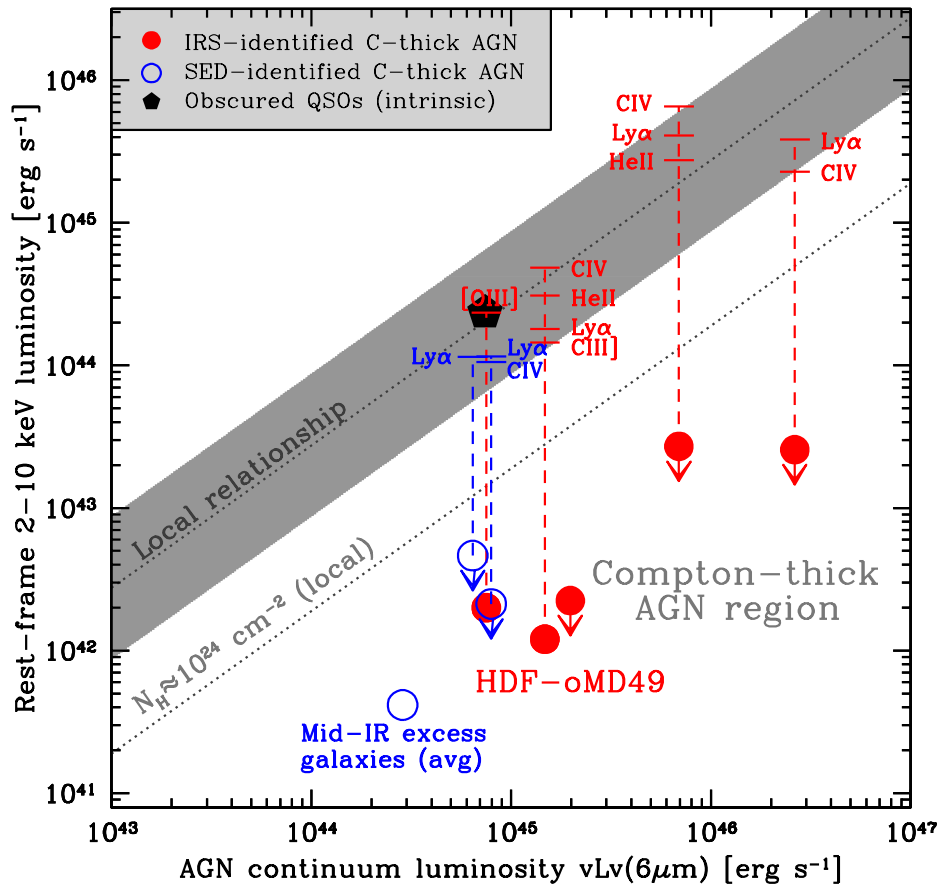}
 \caption{Diagnostic diagram for the identification of Compton-thick AGNs. X-ray luminosity (uncorrected for absorption) is plotted against the 6~$\mu$m AGN luminosity, obtained using either mid-IR spectroscopy or mid-IR spectral energy distributions. The intrinsic (absorption corrected) AGN luminosity is inferred from both the mid-IR AGN component and UV emission lines (as indicated). The plotted AGNs lie at $z\approx$~2 and have mid-IR and emission-line luminosities indicating AGN activity $\approx$~2--3 orders of magnitude higher than indicated by the X-ray luminosity: these are the signatures of a Compton-thick AGN. See Fig.~4 of Alexander et~al. (2008b) for more details.}
 \end{figure}

Significant progress in the identification of X-ray undetected
(potentially Compton thick) AGNs has been made using {\it Spitzer}
observations of $z\approx$~2 massive galaxies (e.g.,\ Daddi
et~al. 2007; Fiore et~al. 2008). For example, selecting galaxies with
excess IR emission over that expected from star-formation activity (on
the basis of absorption-corrected UV emission), Daddi et~al. (2007)
revealed a large population of X-ray undetected IR-luminous galaxies
with either extreme dust-obscured star formation or AGN
activity. Using X-ray stacking analyses, Daddi et~al. (2007) showed
that, although individual galaxies were not detected in the X-ray
band, the average stacked X-ray spectral slope is flat
($\Gamma\approx$~0.9). Such a flat X-ray spectral slope unambiguously
indicates the presence of heavily obscured AGN activity in many
systems, some of which may be obscured by Compton-thick material.

The work of Daddi et~al. (2007) and Fiore et~al. (2008) undoubtably
provide important constraints on the space density of the most heavily
obscured AGNs, and therefore of the most heavily obscured phases of
SMBH growth. However, since these studies were based on X-ray stacking
analyses there are significant uncertainties on (1) the fraction of
obscured AGNs that are contributing to the stacked hard X-ray spectral
slope (e.g.,\ Donley et~al. 2008), (2) the intrinsic luminosities of
the AGNs, and (3) the fraction of the AGNs that are Compton thick,
since both Compton-thin and Compton-thick AGNs can have flat X-ray
spectral slopes. Using the latest 2~Ms {\it Chandra} observations of
the CDF-S (Luo et~al. 2008), we have been able to place constraints on
issues (1) and (2) by detecting individual AGNs that were contributing
to the stack in Daddi et~al. (2007). This new analysis has shown that
the fraction of obscured AGNs contributing to the X-ray stacking
results is at least $\approx$~25\%, and that the AGNs have a broad
range of luminosities from $\approx10^{42}-10^{44}$~erg~s$^{-1}$
(Alexander et~al. 2009). However, without spectroscopic data, none of
these studies are able to reliably determine how many of the AGNs are
Compton thick (issue 3).

The robust identification of distant Compton-thick AGNs requires good
quality spectra in addition to continuum observations. In Alexander
et~al. (2008b) we made great strides in this direction by identifying
luminous AGNs in seven X-ray weak/undetected $z\approx$~2 galaxies
with AGN signatures from optical and/or mid-IR ({\it Spitzer}-IRS)
spectroscopy. The intrinsic X-ray luminosity of the AGNs in these
galaxies could be estimated {\it both} from AGN-dominated optical
emission lines and AGN-dominated mid-IR emission (distinguished from
star formation processes using mid-IR spectroscopy), giving $L_{\rm
X}\approx10^{44}$--$10^{45}$~erg~s$^{-1}$ (i.e.,\ luminous AGN activity). 
By comparison, the observed X-ray luminosities are $\approx$~2--3 orders of
magnitude lower than those predicted, indicating that the X-ray
emission {\it must} be obscured by Compton-thick material; see Fig.~2
and \S3 in Alexander et~al. (2008b). With these diagnostics it is
therefore possible to identify distant Compton-thick AGNs in the
absence of good-quality X-ray spectroscopy. Although limited in source
statistics, the Alexander et~al. (2008b) study was also able to place
basic constraints on the space density of luminous Compton-thick AGNs
($L_{\rm X}>10^{44}$~erg~s$^{-1}$) at $z\approx$~2--2.5. The derived
space density of (0.7--2.5)$\times10^{-5}$~Mpc$^{-3}$ is consistent
with that found for comparably luminous unobscured AGNs, indicating
that Compton-thick SMBH growth was as ubiquitious as unobscured SMBH
growth in the distant Universe. However, since our method relied on the
identification of optical emission lines, these constraints should be
considered a lower limit on the Compton-thick AGN space density as
there are likely to also be many Compton-thick AGNs with weak optical
emission lines (e.g.,\ NGC~6240 in the local Universe).

The Compton-thick AGNs spectroscopically identified in Alexander et~al. 
(2008b) are more luminous than the AGNs studied in Daddi et~al. (2007)
and Fiore et~al. (2008). However, this is clearly a selection effect since in 
order to infer that the X-ray emission from the AGNs in Alexander et~al. 
(2008b) is obscured by Compton-thick material, it was necessary to
select the most luminous AGNs; i.e.,\ given an X-ray sensitivity limit
of $\approx$~10$^{42}$--10$^{43}$~erg~s$^{-1}$,
only objects with intrinsic X-ray luminosities $>10^{44}$~erg~s$^{-1}$
could be reliably identified as Compton thick.
 
\section{Joint Black-Hole--Galaxy Growth in $z\approx$~2 Galaxies}

From a combination of X-ray, IR, and submm data with optical and
mid-IR spectroscopy, we have been able to provide sensitive
constraints on the ubiquity and properties of AGN activity in
$z\approx$~2 dust-obscured galaxies. Can we use these constraints
along with other multi-wavelength data to understand the relative
SMBH--stellar growth in $z\approx$~2 galaxies? Interestingly we can.

The deep multi-wavelength of SMGs show that these $z\approx$~2 galaxies
host both heavily obscured AGN and star formation
activity; see \S2. Qualitatively, this joint SMBH--stellar growth
is consistent with that expected given the
SMBH--spheroid mass relationship in the local Universe. However,
constraints on the masses of the SMBH and spheroid suggest that the
SMBHs in SMGs may be smaller than that expected given the lcoal
SMBH--spheroid mass constraints, indicating that the SMBH may be
``catching up'' with the growth of the SMBH (see Alexander
et~al. 2008a). The estimated growth rate of the SMBHs in SMGs do not
appear to be sufficient to significantly overcome the growth rate of
the spheroid, indicating that a more AGN-dominated growth phase may be
required to place SMGs (and their progeny) on the locally defined
SMBH--spheroid mass relationship (e.g.,\ an optically bright quasar;
Coppin et~al. 2008). Although these constraints suggest that there is
a ``lag'' in the SMBH growth in SMGs, given that SMGs are amongst the
most rapidly evolving galaxies in the Universe, it is perhaps
surprising that the relative SMBH--galaxy spheroid growth rate in
these systems is even {\it approximately} in agreement with that
expected from the SMBH--spheroid mass relationship. This may indicate
that some kind of regulatory mechanism is at work to ``control'' the
growth of these two components (i.e.,\ energetic winds/outflows).

The constraints on the SMBH--galaxy growth in the $z\approx$~2 {\it
Spitzer}-detected galaxies studied by Daddi et~al. (2007) are weaker
than those determined for $z\approx$~2 SMGs since they are typically
less luminous (and therefore fainter). However,
based on the analyses of Daddi et~al. (2007) and Murphy et~al. (2009),
there is clear evidence for both AGN activity and star formation in
many of these systems. By decomposing the IR spectral energy
distributions of the {\it Spitzer}-detected galaxies into AGN and
star-formation components, Daddi et~al. (2007) estimated that
the volume averaged SMBH--stellar growth in these galaxies is
consistent with that expected given the local SMBH--spheroid mass
relationship. The $z\approx$~2 {\it
Spitzer}-detected galaxies are growing less rapidly than the $z\approx$~2 SMGs, and are therefore are
less likely to be able to significantly deviate from the local SMBH--spheroid mass
relationship. However, it is perhaps remarkable that the relative
SMBH--stellar growth rates of both populations are comparable.



\acknowledgements 

I thank the Royal Society and Philip Leverhulme Prize fellowship for
generous support, and my collaborators for allowing me to present this
research.



\end{document}